\documentclass[a4paper,fleqn]{cas-dc}

\usepackage[sort&compress,numbers]{natbib}
\usepackage[utf8]{inputenc}
\usepackage{textcomp}
\DeclareUnicodeCharacter{22EF}{\dots}
\DeclareUnicodeCharacter{0302}{\^}

\usepackage{soul}

\def\tsc#1{\csdef{#1}{\textsc{\lowercase{#1}}\xspace}}
\tsc{WGM}
\tsc{QE}
\tsc{EP}
\tsc{PMS}
\tsc{BEC}
\tsc{DE}

\begin{document}
\let\WriteBookmarks\relax
\def\floatpagepagefraction{1}
\def\textpagefraction{.001}
\shorttitle{Anthraphenylenes: Porous 2D Carbon Monolayers}
\shortauthors{Lima et~al.}

\title [mode = title]{Anthraphenylenes: Porous 2D Carbon Monolayers with Biphenyl-Anthracene Frameworks and Type-II Dirac line nodes}

\author[1,2]{K. A. L. Lima}
\affiliation[1]{
organization={Institute of Physics},
addressline={University of Brasília}, 
city={Brasília },
postcode={70910‑900}, 
state={DF},
country={Brazil}}
\affiliation[2]{
organization={Computational Materials Laboratory, LCCMat, Institute of Physics},
addressline={University of Brasília}, 
city={Brasília },
postcode={70910‑900}, 
state={DF},
country={Brazil}}
\credit{Conceptualization of this study, Methodology, Review and editing, Investigation, Formal analysis, Writing -- review \& editing, Writing -- original draft}

\author[3]{José A. S. Laranjeira}
\cormark[1]
\cortext[cor1]{Corresponding author}
\affiliation[3]{
organization={Modeling and Molecular Simulation Group},
addressline={São Paulo State University (UNESP), School of Sciences}, 
city={Bauru},
postcode={17033-360}, 
state={SP},
country={Brazil}}
\credit{Conceptualization of this study, Methodology, Review and editing, Investigation, Formal analysis, Writing -- review \& editing, Writing -- original draft}
\author[3]{Nicolas F. Martins}
\credit{Conceptualization of this study, Methodology, Review and editing, Investigation, Formal analysis, Writing -- review \& editing, Writing -- original draft}

\author[4]{Sérgio A. Azevedo}
\affiliation[4]{
organization={Federal Institute of Maranhão}, 
city={Barra do Corda},
postcode={65950-000}, 
state={MA},
country={Brazil}}
\credit{Conceptualization of this study, Methodology, Review and editing, Investigation, Formal analysis, Writing -- review \& editing, Writing -- original draft}

\author[3]{Julio R. Sambrano}
\credit{Conceptualization of this study, Methodology, Review and editing, Investigation, Formal analysis, Writing -- review \& editing, Writing -- original draft}

\author[1,2]{L.A. Ribeiro Junior}
\credit{Conceptualization of this study, Methodology, Review and editing, Investigation, Formal analysis, Writing -- review \& editing, Writing -- original draft}

\begin{abstract}
Carbon's versatility allows it to form diverse structures with unique properties, driven by its moderate electronegativity, small ionic radius, and ability to adopt \textit{sp}, \textit{sp\textsuperscript{2}}, and \textit{sp\textsuperscript{3}} hybridizations, individually or in combination. In this work, we introduce three novel 2D carbon allotropes --- $\alpha$, $\beta$, and $\gamma$-anthraphenylenes --- derived from biphenylene and Dewar-anthracene motifs, investigated through density functional theory calculations. Their thermodynamic and dynamic stability are confirmed by cohesive energy ($-7.02$ to $-7.26$ eV/atom), phonon dispersion, and \textit{ab initio} molecular dynamics simulations. The electronic structure analysis shows that all three anthraphenylenes display metallic behavior. All anthraphenylenes feature type-II Dirac Line Nodes (DLNs). Mechanical analysis highlights significant anisotropy, mainly in $\gamma$-anthraphenylene, which exhibits the highest rigidity. These monolayers feature a porous architecture with tunable mechanical properties, making them promising candidates for nanoelectronics and energy storage applications. By expanding the family of 2D carbon materials, anthraphenylenes provide new avenues for functional nanomaterial design.
\end{abstract}




\maketitle

\section{Introduction}

Carbon, a highly versatile element, can form structures in zero (0D), one (1D), two (2D), and three (3D) dimensions, each exhibiting distinct properties \cite{zhao2021versatile,xu2019function,zhang2020multi,song2023versatile}. This adaptability arises from its moderate electronegativity, small atomic radius, and ability to adopt sp, sp\textsuperscript{2}, and sp\textsuperscript{3} hybridizations, within its bonding framework \cite{Kasálková2021Carbon}. Among its various forms, graphene, the most stable 2D allotrope, has attracted significant attention due to its remarkable electronic and spintronic properties, such as massless Dirac fermions and the quantum Hall effect \cite{geim2007rise, bolotin2009observation, taychatanapat2011quantum}. Once these properties are closely tied to its honeycomb lattice, a common strategy for discovering new 2D carbon allotropes involves exploring alternative topologies that may lead to novel electronic and mechanical characteristics.

Several 2D carbon-based materials have been theoretically proposed, many of which exhibit intriguing architectures, out-of-plane geometries, and semiconducting behavior, distinguishing them from graphene \cite{li2017psi, li2021two, lima2025th, girao2023classification, laranjeira2024graphenyldiene, ghorbanali2025tile, niu2023novel,cavalheiro2024can,tiwari2016magical,yang2020screening}. However, defining a practical route for synthesizing these monolayers remains a challenge. Recent advancements in bottom-up approaches have emerged as a promising strategy for obtaining theoretically predicted 2D carbon allotropes \cite{cai2010atomically,bieri2010two,hou2022synthesis,fan2021biphenylene,toh2020synthesis,desyatkin2022scalable,aliev2025planar}. A key aspect of these synthetic methods is the careful selection of molecular precursors and precise control of thermodynamic and kinetic factors to ensure nanoscale precision \cite{palma2011blueprinting,fan2021biphenylene}. For instance, in the synthesis of 2D biphenylene, self-assembled poly(2,5-difluoro-para-phenylene) (PFPP) was employed on an Au(111) substrate \cite{hudspeth2010electronic}. The resulting structure, composed of fused 4–6–8 carbon rings, matched earlier theoretical predictions and was experimentally confirmed to be metallic. Similarly, graphenylene, another biphenylene-based material, was synthesized through polymerization reactions involving 1,3,5-trihydroxybenzene, forming a dodecagonal ring with a diameter of 5.8 \r{A}, closely aligning with computational models \cite{du2017new,fabris2018theoretical}.

In this context, Dewar-anthracene, a metastable isomer of anthracene with a strained bicyclic structure, offers a promising building block for designing novel 2D carbon architectures \cite{PRITSCHINS19821151}. Its synthesis follows a stepwise approach in which the benzocyclobutadiene moiety undergoes a Diels–Alder reaction with an activated dienophile, such as 3,6-dihydrophthalic anhydride, followed by oxidative bis-decarboxylation and controlled dehydrogenation steps \cite{iwata2021synthesis, applequist1964synthesis}. This methodology enables the reversible disruption and reformation of aromaticity, facilitating the construction of extended conjugated systems \cite{PRITSCHINS19821151,iwata2021synthesis}. Dewar-anthracene was recently employed in the design of graphenyldiene, a novel 2D carbon allotrope characterized by a hexagonal lattice composed of fused 4–6–18 membered rings \cite{laranjeira2024graphenyldiene}. This monolayer demonstrated thermal stability up to approximately 1100 K and exhibited semiconducting behavior, highlighting its potential for electronic applications. These findings suggest that integrating biphenylene and Dewar-anthracene motifs could lead to novel 2D carbon materials with tailored electronic and mechanical properties, motivating further exploration of such architectures.

Building on the successful synthesis of biphenylene and graphenylene, as well as the introduction of graphenyldiene, this work presents three novel 2D carbon allotropes --- $\alpha$, $\beta$, and $\gamma$-anthraphenylenes --- derived from biphenylene and Dewar-anthracene motifs. These structures result from slightly expanding the cyclobutadiene rings observed in graphenylene and biphenylene, as illustrated in Figure~\ref{fig:unit_cell}. To characterize their properties, we performed comprehensive density functional theory (DFT) calculations and \textit{ab initio} molecular dynamics (AIMD) simulations, assessing their dynamic and thermal stabilities. We also investigated their electronic structure, optical, and mechanical properties. The results demonstrate that anthraphenylenes enrich the family of 2D carbon materials by combining porous architectures, metallic behavior, infrared and ultraviolet optical absorption, and competitive mechanical properties, positioning them as promising candidates for applications in nanoelectronics and energy storage applications.

\section{Methodology}

DFT calculations \cite{hohenberg1964inhomogeneous} were performed using the Perdew-Burke-Ernzerhof (PBE) \cite{perdew1996generalized} exchange-correlation functional within the Vienna \textit{ab initio} Simulation Package (VASP) \cite{kresse1999ultrasoft}. A plane-wave cutoff energy of 520 eV was set to ensure an accurate representation of electronic states. The Grimme DFT-D2 method was employed to account for van der Waals interactions \cite{grimme2006semiempirical}, while the projected augmented wave (PAW) method \cite{kresse1993ab, kresse1996efficient} was used to describe valence-core electron interactions. A vacuum region of 15 \r{A} was added in the out-of-plane direction to minimize periodic boundary condition artifacts. The Brillouin zone was sampled using the Monkhorst–Pack scheme \cite{monkhorst1976special}, with a $3 \times 3 \times 1$ k-point grid for structural relaxations and an $\Gamma$-centered $18 \times 18 \times 1$ k-point grid for density of states (DOS) calculations.

Structural optimization was carried out using the conjugate gradient algorithm until the convergence criteria were met, ensuring that the total energy variation and atomic forces were below $1 \times 10^{-5}$ eV and 0.01 eV/\r{A}, respectively. The thermal stability of the proposed monolayers was assessed through AIMD simulations in the NVT ensemble, employing the Nos\'e-Hoover thermostat \cite{kresse1993ab,nose1984unified, hoover1985canonical} for 5 ps at 300 K. To verify the dynamic stability, phonon dispersion calculations were performed using the PHONOPY package \cite{togo2015first}, following the finite displacement method.

We have calculated the cohesive energy ($E_{\text{coh}}$) to evaluate the energetic stability of the predicted monolayers relative to other well-established 2D carbon allotropes. It was determined using the expression: $E_{\text{coh}} = (E_{\text{total}} - n_c E_c)/n_c$, where $E_{\text{total}}$ is the total energy of the $\alpha$-, $\beta$-, or $\gamma$-anthraphenylenes, $E_c$ is the energy of an isolated carbon atom, and $n_c$ represents the number of carbon atoms in the unit cell. By this definition, a more negative cohesive energy indicates a more stable structure. 

The electronic properties of the monolayers were investigated by band structure and density of states (DOS) calculations. The band structures were computed along high-symmetry paths in the Brillouin zone, while the DOS was obtained using a denser $\Gamma$-centered $18 \times 18 \times 1$ k-point grid. To gain insight into the charge distribution and chemical bonding characteristics, we analyzed the projected density of states (PDOS) and the electron localization function (ELF) \cite{grin2014elf}. The ELF provides a measure of electron localization, where values close to 1 indicate strong localization, such as in covalent bonds or lone pairs, while values near 0.5 suggest delocalized electron densities, typical of metallic behavior.

We also investigated the mechanical properties of the anthraphenylenes were assessed by computing their elastic constants, including Young's modulus ($Y$), shear modulus ($G$), and Poisson's ratio ($\nu$). The independent elastic constants were obtained using the stress-strain method, where small deformations were applied to the unit cell, and the corresponding stress tensors were calculated. The Born–Huang stability criteria \cite{born1940stability} were used to confirm the mechanical stability of the monolayers. Additionally, the directional dependence of $Y$, $G$, and $\nu$ was analyzed through polar diagrams to evaluate the degree of mechanical anisotropy. 

The optical properties of the anthraphenylenes were examined by calculating the frequency-dependent complex dielectric function, $\varepsilon(\omega) = \varepsilon_1(\omega) + i\varepsilon_2(\omega)$, where $\varepsilon_1(\omega)$ and $\varepsilon_2(\omega)$ represent the real and imaginary components, respectively. The absorption coefficient, $\alpha(\omega)$, and reflectance, $R(\omega)$, were derived from $\varepsilon(\omega)$ using standard relations \cite{moore2020optical}. These properties were computed within the random phase approximation (RPA), neglecting local field effects, to evaluate the response of the monolayers to electromagnetic radiation across different spectral regions.

\section{Results}

\begin{figure*}[pos=htb!]
    \centering
    \includegraphics[width=1\linewidth]{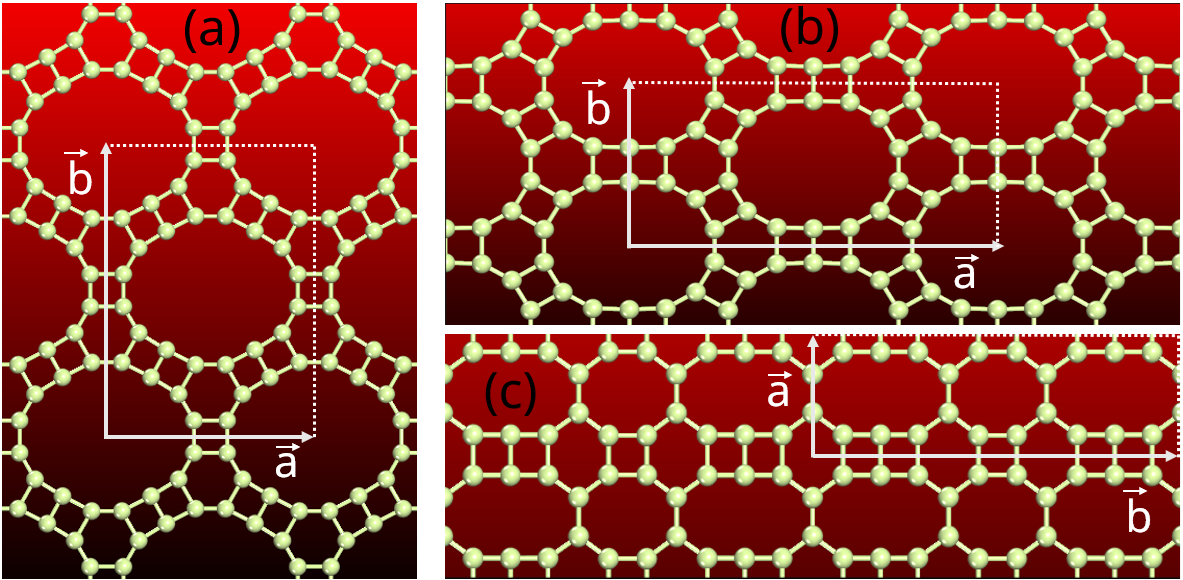}
    \caption{Representative cell with the lattice vectors highlighted for (a) $\alpha$-anthraphenylene, (b) $\beta$-anthraphenylene, and (c) $\gamma$-anthraphenylene.}
    \label{fig:unit_cell}
\end{figure*}

The $\alpha$- and $\beta$-anthraphenylenes are described by rectangular conventional unit cells belonging to the $Cmm$ (No. 65) space group and exhibiting $D_{2h}$ point group symmetry. In contrast, the $\gamma$-anthraphenylene is also represented by a rectangular unit cell but with the $Pmm$ (No. 47) space group while maintaining the $D_{2h}$ point group symmetry. The unit cells of the anthraphenylenes are illustrated in Figure~\ref{fig:unit_cell}. The corresponding lattice parameters are $a = 9.31$ \r{A}, $14.60$ \r{A}, and $4.54$ \r{A}, and $b = 13.14$ \r{A}, $6.76$ \r{A}, and $14.10$ \r{A} for $\alpha$-, $\beta$-, and $\gamma$-anthraphenylenes, respectively. These monolayers consist of carbon-fused rings arranged in different patterns: $\alpha$-anthraphenylene features a 4–6–16 ring configuration, $\beta$-anthraphenylene has a 4–6–14 configuration, and $\gamma$-anthraphenylene consists of a 4–6–8–10 arrangement. The maximum pore diameters of these structures are 7.96 \r{A}, 6.50 \r{A}, and 5.38 \r{A} for $\alpha$-, $\beta$-, and $\gamma$-anthraphenylenes, respectively. Table~\ref{tab:properties_comparison} summarizes the lattice parameters, space groups, band gap energies, and cohesive energies of the anthraphenylenes alongside other 2D carbon allotropes. For the sake of comparison, these properties for the other allotropes were also calculated in this work.

\begin{table*}[pos=htb!]
\centering
\caption{Lattice parameters ($a$ and $b$ in Å, $\alpha = \beta$, and $\gamma$ in $^\circ$), space group (SPG), band gap energy ($E_\text{gap}$) in eV, and cohesive energy ($E_\text{coh}$) in eV/atom obtained for $\alpha$, $\beta$, and $\gamma$-anthraphenylenes and other relevant carbon monolayers calculated in this work at DFT/GGA-PBE level.}
\label{tab:properties_comparison} 
\begin{tabular}{llllllll}
\hline
  & $a$ & $b$ & $\alpha = \beta$ & $\gamma$ & SPG & $E_\text{gap}$ & $E_\text{coh}$ \\
\hline
$\alpha$-anthraphenylene   & 9.31 & 13.14 & 90 & 90  & $Cmm$ (65)    & Conductor      & -7.02  \\
$\beta$-anthraphenylene   & 14.60 & 6.76  & 90 & 90  & $Cmm$ (65)    & Conductor      & -7.15  \\
$\gamma$-anthraphenylene       & 4.54 & 14.10  & 90 & 90  & $Pmmm$ (47)    & Conductor & -7.26  \\
T-graphene  \cite{sheng2011t}      & 3.45 & 3.45  & 90 & 90 & $P4/mmm$ (123) & Conductor      & -7.45  \\
Twin-graphene \cite{jiang2017twin}    & 6.14 & 6.14  & 90 & 120 & $P6/mmm$ (191) & 0.73 & -7.08  \\
PHE-graphene \cite{zeng2019new}  & 5.73 & 5.73  & 90 & 120 & $P\overline{6}m2$ (187)   & Conductor      & -7.56  \\
Penta-graphene \cite{zhang2015penta} & 3.64 & 3.64 & 90 & 90 & $P42_1m$ (113) & 2.19 & -7.13 \\
Graphyine \cite{narita1998optimized} & 9.46  & 9.46  & 90 & 120 & $P6/mmm$ (191)    & Conductor & -7.20 \\
Graphenyldiene  \cite{laranjeira2024graphenyldiene}   & 6.07 & 6.07  & 90 & 90  & $P4/mbm$ (127)  & 0.78      & -6.92  \\
Graphenylene \cite{song2013graphenylene, du2017new}    & 6.77 & 6.77  & 90 & 120  & $P6/mmm$ (191)  & 0.04      & -7.33  \\
Biphenylene \cite{fan2021biphenylene} & 3.80 & 4.50 &  90 & 90 & $Pmmm$ (47) & Conductor & -7.47 \\
Graphene  \cite{geim2007rise}   & 2.47 & 2.47  & 90 & 120  & $P6/mmm$ (191)  & Conductor      & -8.02  \\
\hline
\end{tabular}
\end{table*}

The cohesive energy ($E_{\text{coh}}$) values of the $\alpha$-, $\beta$-, and $\gamma$-anthraphenylenes were calculated to assess their thermodynamic stability. The obtained values of $-7.02$, $-7.15$, and $-7.26$ eV/atom, respectively, are comparable to those of well-known 2D carbon allotropes such as twin-graphene ($-7.08$ eV/atom) and penta-graphene ($-7.13$ eV/atom). These results indicate that anthraphenylenes are energetically favorable structures. $\gamma$-anthraphenylene exhibits the lowest cohesive energy among the three, suggesting it is the most stable configuration. This trend in $E_{\text{coh}}$ can be attributed to variations in the monolayers' bonding arrangements and strain distributions. Overall, the cohesive energy analysis confirms that anthraphenylenes are competitive in stability compared to existing 2D carbon allotropes.

\begin{figure*}[pos=htb!]
    \centering
    \includegraphics[width=1\linewidth]{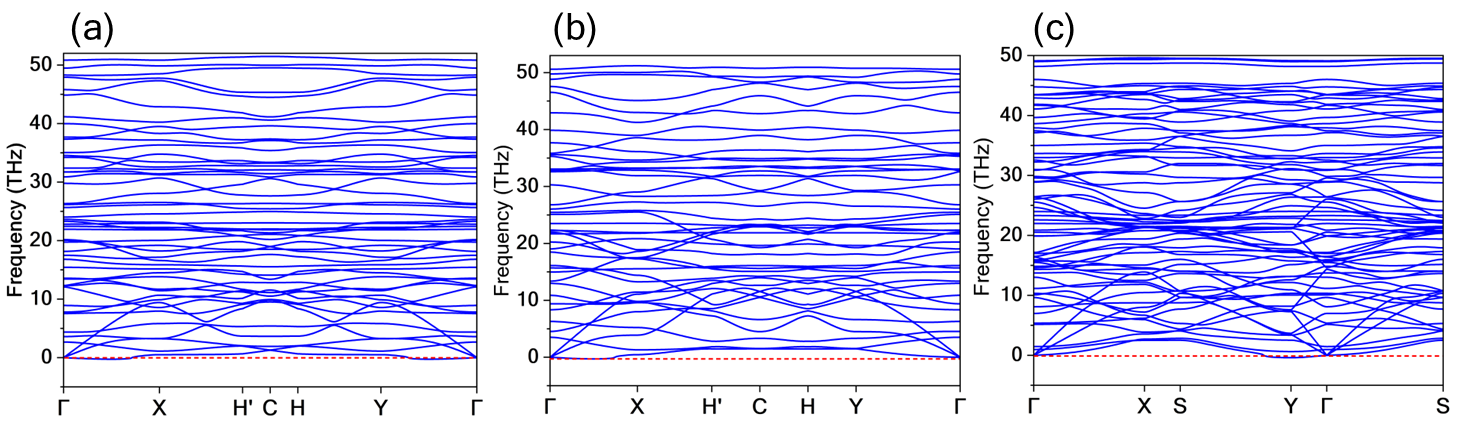}
    \caption{Phonon dispersion calculated along the high-symmetry Brillouin zone paths for (a) $\alpha$-anthraphenylene, (b) $\beta$-anthraphenylene, and (c) $\gamma$-anthraphenylene.}
    \label{fig:phonon}
\end{figure*}

To evaluate the dynamic stability of the anthraphenylenes, phonon dispersion calculations were performed along the high-symmetry paths of the Brillouin zone, as shown in Figure~\ref{fig:phonon}. The results reveal that all monolayers exhibit vibrational frequencies extending up to approximately 50 THz. Although slight imaginary frequencies are observed at the $\Gamma$ point in all cases, they are minimal. The most significant negative frequency is found in $\gamma$-anthraphenylene, reaching approximately $-0.39$ THz ($-13.25$ cm$^{-1}$), while $\alpha$- and $\beta$-anthraphenylenes exhibit lower values of $-0.05$ THz ($-1.67$ cm$^{-1}$) and $-0.13$ THz ($-4.40$ cm$^{-1}$), respectively. These small imaginary modes may arise from finite-size effects in the supercell calculations or intrinsic stress. However, given that a threshold of approximately $-2$ THz is often used to assess the stability of free-standing 2D materials \cite{wang2022high}, the observed frequencies are within an acceptable range, supporting the dynamic stability of anthraphenylenes.

\begin{figure*}[pos=htb!]
    \centering
    \includegraphics[width=1\linewidth]{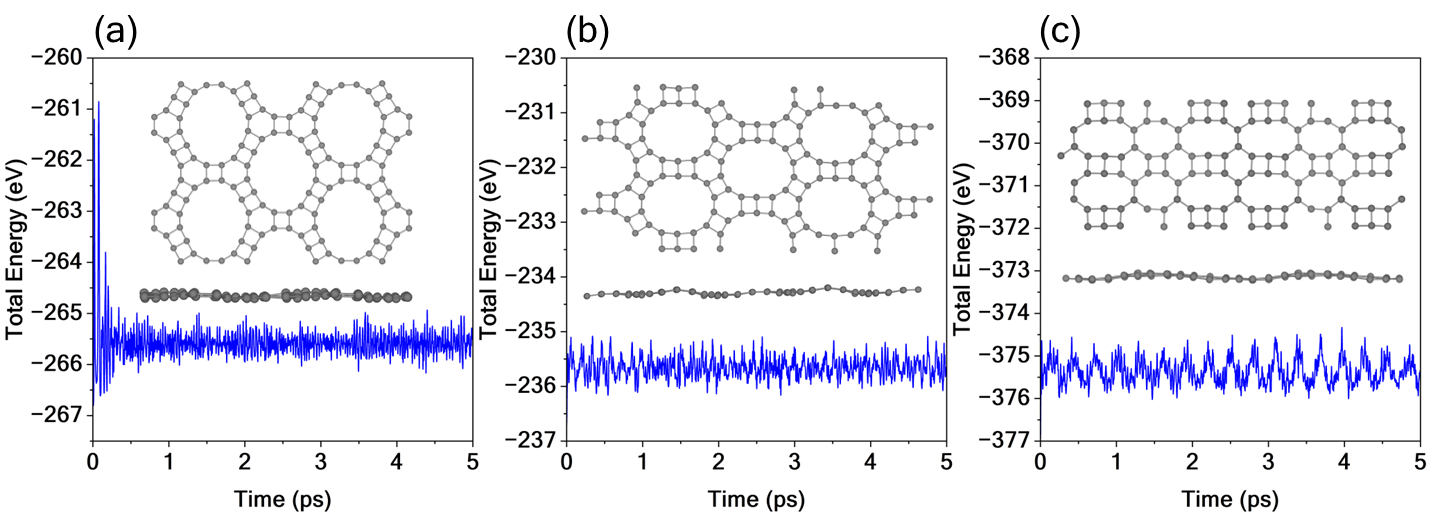}
    \caption{\textit{ab initio} molecular dynamics simulations conducted at 300 K during 5 ps for (a) $\alpha$-anthraphenylene, (b) $\beta$-anthraphenylene, and (c) $\gamma$-anthraphenylene.}
    \label{fig:aimd}
\end{figure*}

The thermal stability of the anthraphenylenes was further examined through AIMD simulations at 300 K for 5 ps, as shown in Figure~\ref{fig:aimd}. The total energy fluctuations remained minimal throughout the simulations, indicating that all three monolayers retain their structural integrity at room temperature. No significant atomic rearrangements or bond breakages were observed, confirming their stability. The slight variations in energy can be attributed to thermal vibrations, which are expected in finite-temperature simulations. Combined with the phonon dispersion analysis, these results strongly suggest that $\alpha$-, $\beta$-, and $\gamma$-anthraphenylenes are dynamically and thermally stable under ambient conditions.

\begin{figure*}[pos=htb!]
    \centering
    \includegraphics[width=1\linewidth]{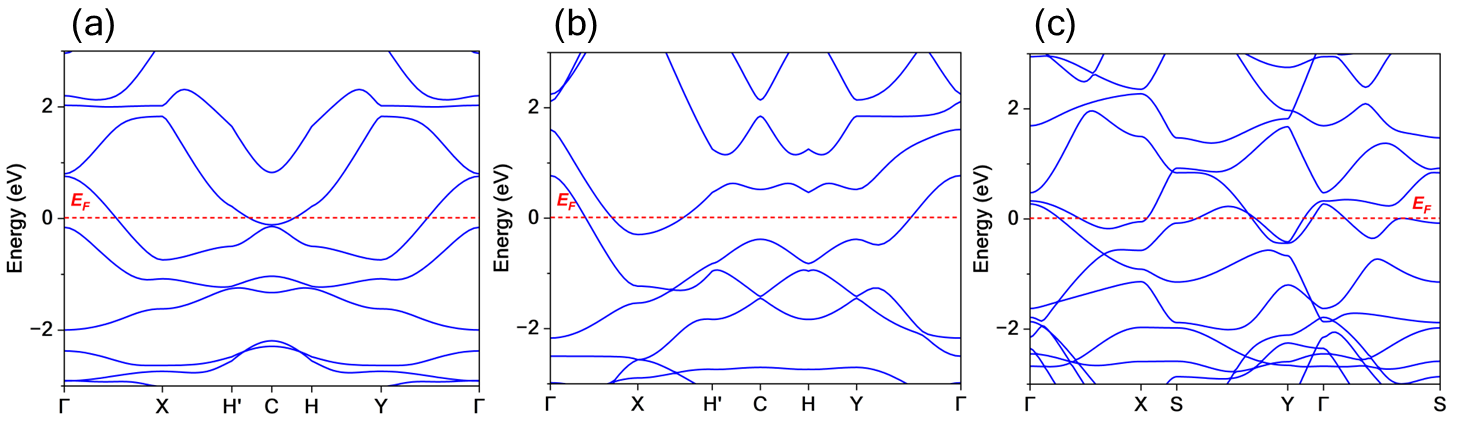}
    \caption{Band structure for (a) $\alpha$-anthraphenylene, (b) $\beta$-anthraphenylene, and (c) $\gamma$-anthraphenylene with the Fermi level in red dashed lines.}
    \label{fig:bands}
\end{figure*}

\begin{figure*}[pos=htb!]
    \centering
    \includegraphics[width=1\linewidth]{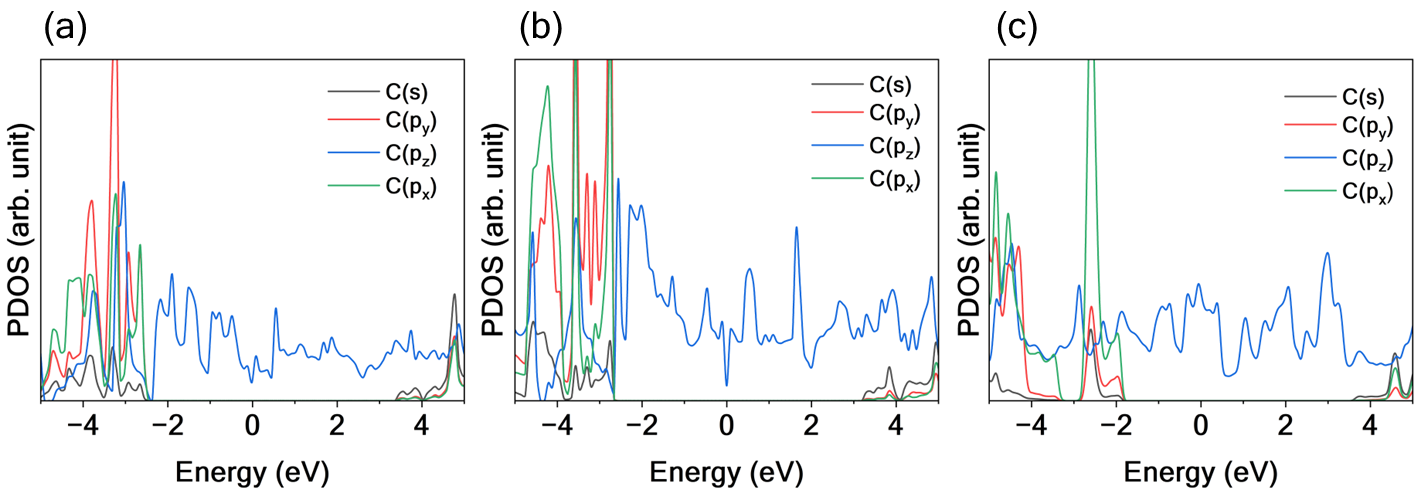}
    \caption{Projected density of states (PDOS) for (a) $\alpha$-anthraphenylene, (b) $\beta$-anthraphenylene, and (c) $\gamma$-anthraphenylene.}
    \label{fig:pdos}
\end{figure*}

The electronic properties of the anthraphenylenes were investigated through band structure and projected density of states (PDOS) calculations, as presented in Figure~\ref{fig:bands} and~\ref{fig:pdos}. The results indicate that all three monolayers exhibit metallic behavior, with electronic bands crossing the Fermi level. For both $\alpha$-anthraphenylene and $\beta$-anthraphenylene, the bands near the Fermi level ($E_F$) exhibit greater dispersion, around 2 eV, which is reflected in the projected density of states (PDOS), where a noticeable decrease in the density of states (DOS) at 0 eV is observed. These structures also feature type-II Dirac Line Nodes (DLNs), formed by interactions between the two highest valence bands. These DLNs correspond to topological states in which energy bands intersect along continuous lines in momentum space, enabling unique electronic transport phenomena. Specifically, for $\alpha$-anthraphenylene, the DLNs appear at the H' and H points, while for $\beta$-anthraphenylene, they are found at the paths \(X\ \rightarrow H'\), and \(Y\ \rightarrow \Gamma\), and are more evident at $C$ and $Y$ points.

In contrast, $\gamma$-anthraphenylene shows a higher concentration of electronic bands around $E_F$ compared to $\alpha$ and $\beta$-anthraphenylenes, consistent with the DOS, which remains elevated in this region. In addition, type-II DLNs are identified in the directions \(\Gamma\ \rightarrow X\), and \(Y\ \rightarrow \Gamma\). 

The projected density of states (PDOS) analysis, shown in Figure~\ref{fig:pdos}, reveals that the metallic nature of these monolayers is primarily governed by $p_z$ orbitals, highlighting the significant $\pi$ character of their electronic states. At lower valence band (VB) energies, noticeable contributions from the $s$, $p_x$, and $p_y$ orbitals emerge. Specifically, $\gamma$-anthraphenylene exhibits an enhanced density of states around $-2$ eV, indicating more pronounced sp$^2$ hybridization, as reflected in the similar shapes of the PDOS curves in this region. In contrast, $\alpha$- and $\beta$-anthraphenylenes show a moderate degree of sp$^2$ hybridization, with contributions from $s$, $p_x$, and $p_y$ orbitals in the same energy range, albeit with distinct PDOS profiles.

\begin{figure*}[pos=htb!]
    \centering
    \includegraphics[width=1\linewidth]{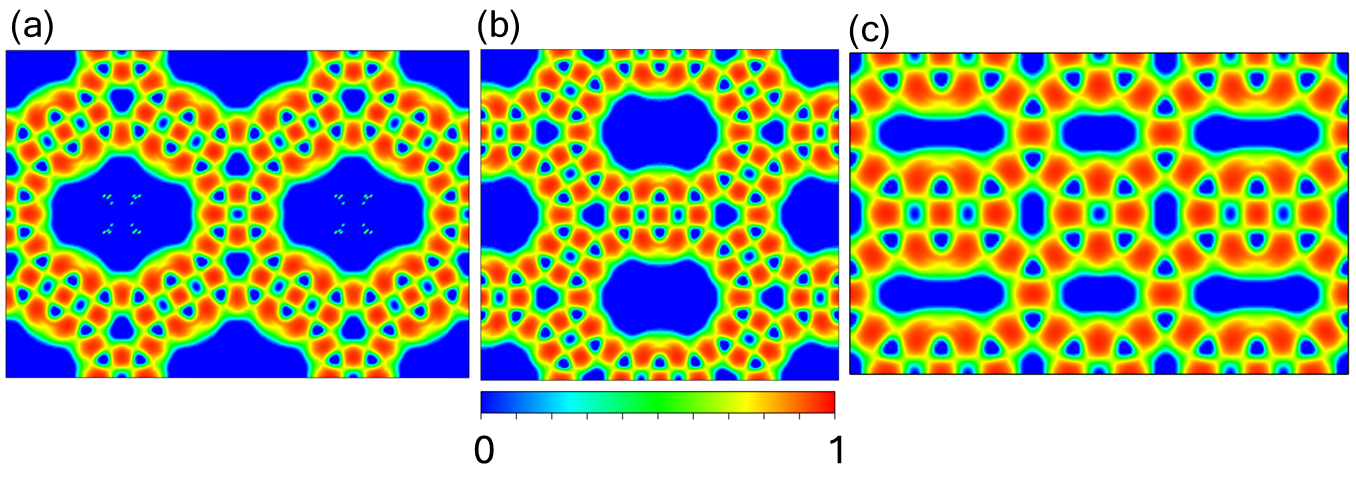}
    \caption{Electron Function Localization (ELF) for (a) $\alpha$-anthraphenylene, (b) $\beta$-anthraphenylene, and (c) $\gamma$-anthraphenylene.}
    \label{fig:ELF}
\end{figure*}

To gain a deeper understanding of the charge distribution and bonding characteristics, the electron localization function (ELF) was analyzed, as shown in Figure~\ref{fig:ELF}. An ELF value of 1 corresponds to perfect electron localization, typically found in covalent bonds or lone pairs, where electrons are tightly bound. Conversely, a value of 0.5 indicates a more delocalized electron distribution, resembling a uniform electron gas, while an ELF of 0 points to regions with little or no electron density. In this way, the ELF maps reveal significant electron localization along the bond axes of benzene rings, consistent with the resonant $\pi$-system characteristic of aromatic compounds. 

Despite the planarity of the Dewar-benzene rings, which enhances orbital overlap compared to nonplanar counterparts, the ELF indicates weaker electron localization along the bonds linking the cyclobutadiene units. This feature suggests that while conjugation is present, it is less effective than benzene due to inherent bond strain and altered electron distribution. Additionally, the biphenyl groups exhibit high electron localization along their cyclobutadiene rings, reflecting the overlap of $p$-orbitals. However, the reduced resonance stabilization compared to benzene highlights the delicate interplay between planarity, bond strain, and $\pi$-electron delocalization in anthraphenylenes.

\begin{figure*}[pos=htb!]
    \centering
    \includegraphics[width=1\linewidth]{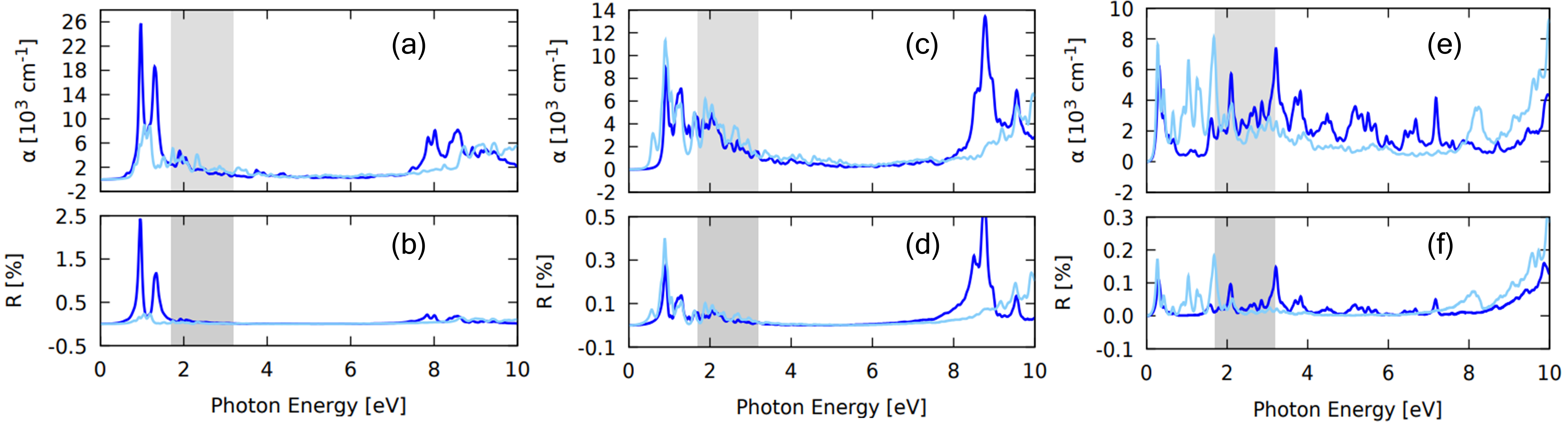}
    \caption{Absorption spectrum and Reflectance calculated for (a) and (b) $\alpha$-anthraphenylene, (c) and (d) $\beta$-anthraphenylene, and (e) and (f) $\gamma$-anthraphenylene. The gray area indicates the visible region of the spectrum. The dark and light blue colors refer to the light polarization along the x- and y-direction, respectively.}
    \label{fig:optical}
\end{figure*}

The optical properties of the anthraphenylenes were examined through the absorption coefficient $\alpha$ and reflectance $R$ spectra as functions of photon energy, as shown in Figure~\ref{fig:optical}. In this figure, the gray area indicates the visible region of the spectrum. All three monolayers exhibit relatively low absorption in the infrared region, indicating a high degree of transparency to long-wavelength radiation. In the visible range, distinct absorption features emerge. $\alpha$-anthraphenylene displays pronounced absorption bands starting around 1.5 eV, corresponding to near-infrared and red wavelengths. The reflectance spectrum follows a similar trend, peaking at approximately 2.5\%, consistent with strong absorption. $\beta$-anthraphenylene, in contrast, exhibits a slightly red-shifted absorption profile, with the onset of strong absorption around 1.0 eV, suggesting optical transparency under standard lighting conditions while absorbing in the infrared range. $\gamma$-anthraphenylene presents a more complex optical response, with absorption peaks spanning a broader energy range. Although it exhibits some absorption in the visible spectrum, its most intense transitions occur in the ultraviolet region beyond 8 eV. The reflectance, while generally low, increases gradually at higher photon energies.

\begin{figure*}[pos=htb!]
    \centering
    \includegraphics[width=1\linewidth]{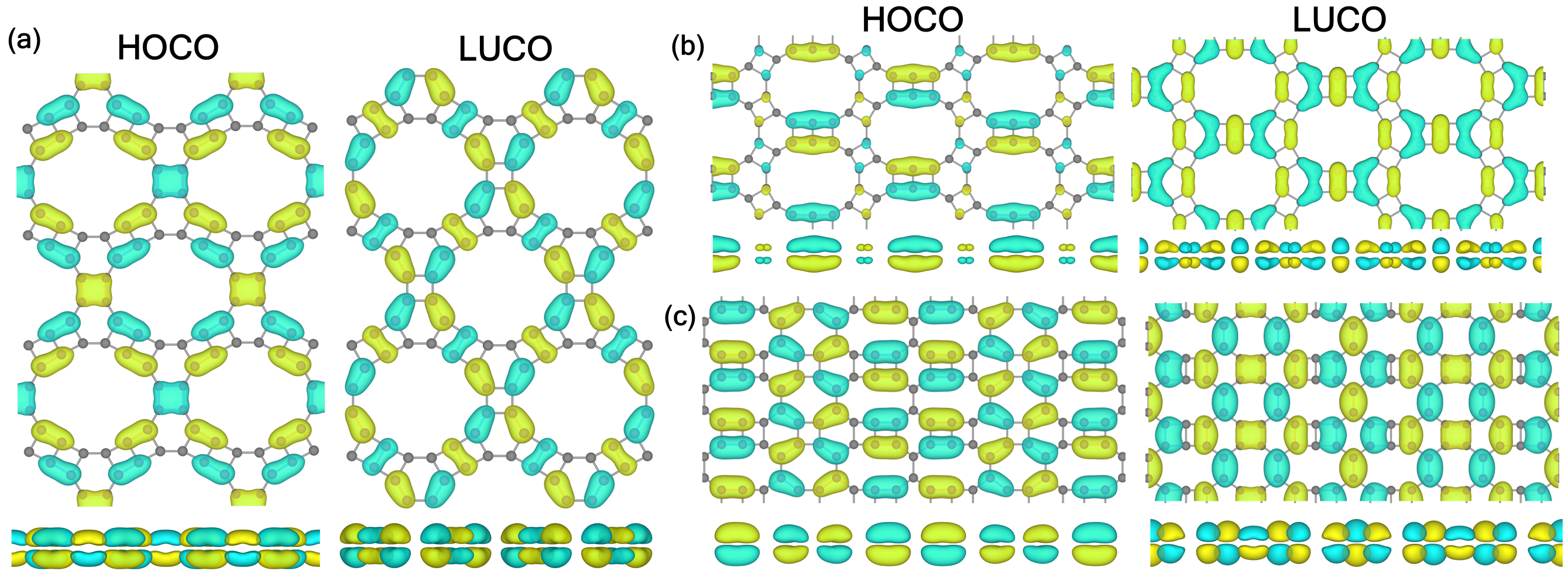}
    \caption{Highest Occupied Crystalline Orbital (HOCO) and Lowest Unoccupied Crystalline Orbital (LUCO) for (a) $\alpha$-anthraphenylene, (b) $\beta$-anthraphenylene, and (c) $\gamma$-anthraphenylene.}
    \label{fig:HOCO_LUCO}
\end{figure*}

Now, we turn to the charge density distributions of the highest occupied crystalline orbital (HOCO) and lowest unoccupied crystalline orbital (LUCO), which were analyzed to explore the electronic properties of anthraphenylenes further, as shown in Figure~\ref{fig:HOCO_LUCO}. For $\alpha$-anthraphenylene, the HOCO is primarily localized on the cyclobutadiene rings of the biphenyl fragments and the bonds within the Dewar-benzene rings, while the LUCO states extend over the benzene rings and the connecting bonds between cyclobutadienes in the Dewar-benzene units.

A similar distribution is observed in $\beta$-anthraphenylene, where the HOCO states are mainly associated with the Dewar-benzene rings and cyclobutadiene units of the biphenyl groups, while the LUCO states are delocalized across the benzene rings and linking bonds. In both cases, the HOCO and LUCO orbitals share localization patterns, reinforcing the presence of strong $\pi$-electron delocalization. As highlighted in the ELF analysis, the Dewar-benzene rings exhibit the highest degree of electron delocalization, reinforcing their pronounced $\pi$-character.

In contrast, $\gamma$-anthraphenylene exhibits a more distinct distribution, with the HOCO concentrated in the Dewar-benzene and cyclobutadiene rings of the biphenyl groups, while the LUCO states appear not only in the Dewar rings and their adjacent benzene rings but also in the cyclobutadiene rings and the bonds linking biphenyl and Dewar-benzene units. Importantly, these results confirm the strong $\pi$-character of the electronic states in anthraphenylenes, contributing to their metallic nature and potential for high carrier mobility.

The mechanical properties of anthraphenylenes were evaluated by computing their elastic constants, including Young’s modulus ($Y$), shear modulus ($G$), and Poisson’s ratio ($\nu$). We have examined and compared the obtained results regarding other relevant 2D carbon structures such as graphene, graphenyldiene, and graphyne (see Table~\ref{tab:elastic_constants}). The calculated elastic constants confirm that all three monolayers satisfy the Born–Huang stability criteria ($C_{11}$ > 0, $C_{66}$ > 0, and $C_{11}$$C_{22}$ > $C_{12}$$C_{12}$) \cite{born1940stability}, ensuring their mechanical stability. Four nonnull elastic constants were reported $C_{11}$ = 186.72 N/m, 230.74 N/m, 185.37 N/m, $C_{12}$ = 59.15 N/m, 53.01 N/m, 77.53 N/m, $C_{22}$ = 146.62 N/m, 184.45 N/m, 313.42 N/m and $C_{66}$ = 65.29 N/m, 64.35 N/m, 57.35 N/m for $\alpha$, $\beta$ and $\gamma$ analogues, respectively.  

\begin{table*}[pos=!htb]
\centering
\caption{Maximum and minimum values of Young's modulus (\(Y_{\text{max}}, Y_{\text{min}}\)) (N/m), Poisson ratio (\(\nu_{\text{max}}, \nu_{\text{min}}\)), and Shear modulus (\(G_{\text{max}}, G_{\text{min}}\)) (N/m) for $\alpha$, $\beta$, and $\gamma$-anthraphenylenes and other carbon monolayers.}
\begin{tabular}{lccc}
\hline
 & \(Y_{\text{max}}/Y_{\text{min}}\) & \(\nu_{\text{max}}/\nu_{\text{min}}\) & \(G_{\text{max}}/G_{\text{min}}\) \\
\hline
$\alpha$-anthraphenylene & 169.94/127.89 & 0.40/0.26 & 65.29/52.87 \\ 
$\beta$-anthraphenylene & 215.50/167.53 & 0.34/0.23 & 76.26/64.35 \\
$\gamma$-anthraphenylene & 281.00/158.27 & 0.47/0.25 & 79.67/57.35 \\
PHE-graphene & 262.29/262.29 & 0.26/0.26 & 103.91/103.91 \\
C3-9H & 238.80/238.80 & 0.26/0.26 & 94.66/94.66 \\
Graphenylene & 209.02/209.02 & 0.27/0.27 & 82.11/82.11 \\
Graphene & 345.42/345.42 & 0.17/0.17 & 147.60/147.60 \\
Graphenyldiene & 122.47/122.47 & 0.35/0.35 & 45.29/45.29 \\
Penta-graphene & 271.81/266.67 & -0.08/-0.10 & 151.21/144.98 \\
T-graphene & 293.90/148.02 & 0.16/0.58 & 126.57/148.02 \\ 
Graphyne & 123.59/123.59 & 0.45/0.45 & 42.63/42.63 \\
\hline
\end{tabular}
\label{tab:elastic_constants}
\end{table*}

The maximum Young's modulus values ($Y_{\text{max}}$) are 169.94 N/m, 215.50 N/m, and 281.00 N/m for $\alpha$-, $\beta$-, and $\gamma$-anthraphenylenes, respectively, indicating that the $\gamma$ variant is the stiffest structure. The Poisson’s ratio varies from 0.23 for $\beta$-anthraphenylene to 0.47 for $\gamma$-anthraphenylene, suggesting significant differences in lateral deformation behavior. Similarly, the shear modulus follows a comparable trend, with $\gamma$-anthraphenylene exhibiting the highest resistance to shear deformation. These findings highlight the mechanical robustness of anthraphenylenes and their potential suitability for applications requiring mechanically resilient 2D materials.

\begin{figure*}[pos=htb!]
    \centering
    \includegraphics[width=1\linewidth]{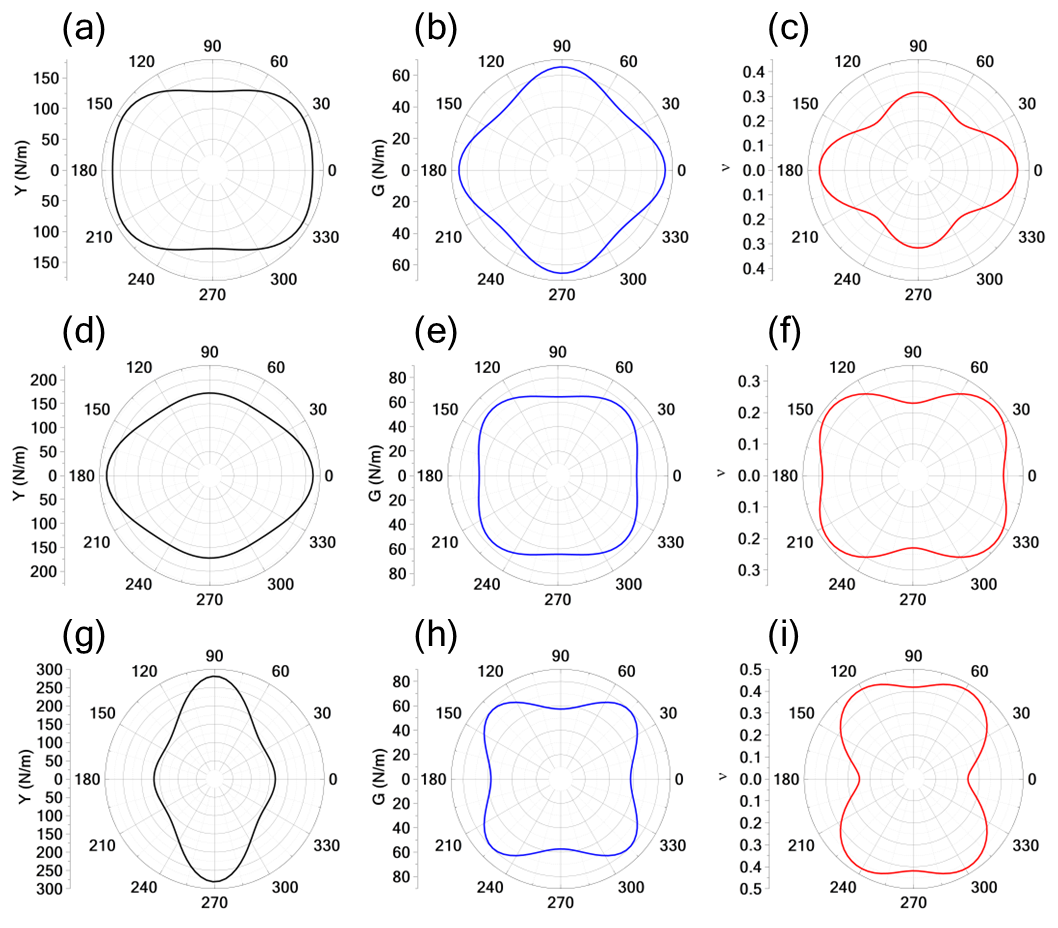}
    \caption{Polar representations of Young's modulus (Y), shear modulus (G), and Poisson's ratio ($\nu$) for (a, b, and c) $\alpha$-anthraphenylene, (d, e, and f) $\beta$-anthraphenylene, and (g, h, and i) $\gamma$-anthraphenylene.}
    \label{fig:polar}
\end{figure*}

In Figure~\ref{fig:polar}, the mechanical anisotropy of anthraphenylenes was further analyzed by examining the directional dependence of Y, G, and $\nu$ through polar diagrams. Among the three structures, $\gamma$-anthraphenylene exhibits the most pronounced anisotropy, with Y values ranging from 158.27 N/m to 281.00 N/m and a Poisson's ratio varying between 0.25 and 0.47. This significant variation suggests that its mechanical response is highly direction-dependent. In contrast, $\beta$-anthraphenylene displays moderate anisotropy, with Young's modulus values spanning from 167.53 N/m to 215.50 N/m and a Poisson's ratio ranging from 0.23 to 0.34, indicating a more uniform stiffness distribution. $\alpha$-anthraphenylene, while exhibiting the lowest overall stiffness (127.89 N/m to 169.94 N/m), demonstrates a relatively isotropic response, as reflected in the symmetric nature of its polar plots. Compared to other 2D carbon allotropes, such as graphene (345.42 N/m), which is nearly isotropic due to its symmetric honeycomb lattice, anthraphenylenes display notable anisotropic mechanical properties. This directional dependence suggests potential applications where tunable mechanical responses are advantageous, such as in flexible electronics or strain-engineered nanodevices.

Unlike other 2D carbon allotropes, such as graphene, graphenylene, and PHE-graphene, which exhibit nearly isotropic mechanical behavior, anthraphenylenes demonstrate significant mechanical anisotropy. This deviation arises from their unique structural motifs, composed of biphenyl and Dewar-anthracene units, which introduce directional variations in bond stiffness and strain distribution. The pronounced anisotropy observed in $\gamma$-anthraphenylene suggests that it may be particularly well-suited for applications requiring materials with tunable mechanical responses. For instance, strain-engineered nanodevices and flexible electronic components could benefit from their direction-dependent stiffness and deformation characteristics. These findings highlight the potential of anthraphenylenes as mechanically versatile 2D materials, broadening the scope of their applicability in emerging technologies.

Finally, the shear modulus, which measures a material's resistance to shear stress, follows a similar trend to Young's modulus. $\gamma$-anthraphenylene exhibits the highest G value at 79.67 N/m, followed by $\beta$-anthraphenylene at 76.26 N/m. These values are significantly higher than that of graphenyldiene (45.29 N/m) but remain lower than graphene (147.60 N/m), highlighting that while anthraphenylenes provide moderate shear resistance, graphene remains superior in resisting shear deformation. Among the anthraphenylenes, the $\gamma$ variant stands out as the most mechanically robust, making it particularly suitable for applications requiring both flexibility and resistance to shear stress.

\section{Conclusion}

In summary, we introduced three novel 2D carbon allotropes --- $\alpha$-, $\beta$-, and $\gamma$-anthraphenylenes --- derived from biphenylene and Dewar-anthracene motifs. Structural stability was confirmed through cohesive energy calculations, phonon dispersion analysis, and AIMD simulations. The cohesive energy values of $-7.02$, $-7.15$, and $-7.26$ eV/atom for $\alpha$-, $\beta$-, and $\gamma$ variants, respectively, indicate their thermodynamic favorability, with $\gamma$-anthraphenylene emerging as the most stable configuration. Phonon dispersion results showed only minor imaginary frequencies, while AIMD simulations demonstrated that all three monolayers maintain structural integrity at 300 K, confirming their dynamic and thermal stability.

The electronic structures of $\alpha$- and $\beta$-anthraphenylenes reveal type-II Dirac Line Nodes (DLNs) near $E_F$, with bands displaying $\sim$2 eV dispersion and a decrease in the density of states (DOS) at 0 eV. For $\alpha$-anthraphenylene, the DLNs appear at the H' and H points, while for $\beta$-anthraphenylene, they are more evident along the $X \rightarrow H'$ and $Y \rightarrow \Gamma$ paths, particularly at the $C$ and $Y$ points. In contrast, $\gamma$-anthraphenylene shows a higher concentration of bands around $E_F$, aligning with an elevated DOS, and features type-II DLNs along the $\Gamma \rightarrow X$ and $Y \rightarrow \Gamma$ directions.

The PDOS analysis confirmed that the metallic nature of these materials is primarily governed by $p_z$ orbitals, emphasizing the strong $\pi$-character of their electronic states. Additionally, the ELF analysis demonstrated significant electron delocalization in the Dewar-benzene rings, further reinforcing the role of $\pi$-electron interactions in shaping the electronic properties of these monolayers.

The mechanical properties of anthraphenylenes were also investigated, revealing significant anisotropy. The $\gamma$ variant exhibited the highest Young’s modulus (281.00 N/m) and shear modulus (79.67 N/m), indicating superior stiffness and resistance to shear deformation. In contrast, $\alpha$-anthraphenylene demonstrated the lowest stiffness, while $\beta$-anthraphenylene displayed intermediate values. The Poisson’s ratio varied between 0.23 and 0.47, suggesting distinct lateral deformation behaviors among the three monolayers. These findings highlight the mechanical robustness of anthraphenylenes, with $\gamma$-anthraphenylene emerging as an up-and-coming candidate for applications requiring both flexibility and structural integrity.

\section*{Data access statement}
Data supporting the results can be accessed by contacting the corresponding author.

\section*{Conflicts of interest}
The authors declare no conflict of interest.

\section*{Acknowledgements}
This work was supported by the Brazilian funding agencies Fundação de Amparo à Pesquisa do Estado de São Paulo - FAPESP (grant no. 2022/03959-6, 2022/00349- 2, 2022/14576-0, 2020/01144-0, 2024/05087-1, and 2022/16509-9), and National Council for Scientific, Technological Development - CNPq (grant no. 307213/2021–8). L.A.R.J. acknowledges the financial support from FAP-DF grants 00193.00001808/2022-71 and $00193-00001857/2023-95$, FAPDF-PRONEM grant 00193.00001247/2021-20, PDPG-FAPDF-CAPES Centro-Oeste 00193-00000867/2024-94, and CNPq grants $350176/2022-1$ and $167745/2023-9$. 

\printcredits

\bibliography{biblio/cas-refs}

\end{document}